\documentclass{iaus}
\usepackage{graphicx}
\usepackage[authoryear]{natbib}

\title[ZDI of II Peg] 
{Zeeman-Doppler Imaging of II Peg }

\author[Carroll et al.]   
{T.A. Carroll$^1$, M. Kopf$^1$, K.G. Strassmeier$^1$, I. Ilyin$^1$ and I. Tuominen$^2$}

\affiliation{$^1$Astrophysikalisches Institut Potsdam,
An der Sternwarte 16, D-14882 Potsdam, Germany \break email: tcarroll@aip.de \\[\affilskip]
$^2$ University of Helsinki, PO Box 14, 00014 Helsinki, Finland}

\pubyear{2009}
\volume{259}  
\pagerange{119--126}
\date{?? and in revised form ??}
\jname{Proceedings Title IAU Symposium}
\editors{A.C. Editor, B.D. Editor \& C.E. Editor, eds.}
\begin{document}

\maketitle

\begin{abstract}
We present Zeeman-Doppler images of the active K2 star II Peg for the years 2004 and
2007. The surface  magnetic field was reconstructed with our new ZDI code \emph{iMap}
which provides a full polarized radiative transfer driven inversion to simultaneously
reconstruct the surface temperature and magnetic vector field distribution. II Peg
shows a remarkable large scale magnetic field structure for both years. 
The magnetic field is predominantly located at high latitudes and is arranged 
in active longitudes. A dramatic evolution in the 
magnetic field structure is visible for the two years, where a 
dominant and largely unipolar field in 2004 has developed into two distinct 
and large scale bipolar structures in 2007.  

\keywords{stars: activity, stars: magnetic fields, stars: spots, radiative transfer, 
methods: data analysis}
\end{abstract}

\firstsection 
\section{Introduction}

Zeeman-Doppler imaging (ZDI) also known as Magnetic-Doppler imaging (MDI) is a powerful inversion 
method to reconstruct stellar magnetic surface fields and has significantly contributed to our current
understanding of stellar magnetic fields \citep[e.g.][]{Donati97,Donati99,
Oleg04,Donati06,Donati07}.
As ZDI is a non-linear inverse problem it critically depends on the underlying forward modeling
and particular attention must be given to a correct modeling of radiative transfer effects 
(see Carroll et al. this proceeding).
In an effort to pursue a more rigorous modeling we have developed our ZDI code \emph{iMap}
which incorporates a full polarized radiative transfer based inversion \citep{Carroll07,Carroll08}. 
The inverse module is based on either a conjugate gradient or Levenberg-Marquardt method. 
The regularization that we apply is based on a new local maximum entropy \citep{Carroll07}. 
One of the great obstacles for a radiative transfer based interpretation 
of observed Stokes profiles is the noise level.
Although there exist a powerful multi-line reconstruction technique, i.e. the least-square
deconvolution method of \cite{Donati97}, the interpretation of the reconstructed \emph{mean} line profiles
is not straight forward.
Another method which provides a way of reconstructing individual Stokes line
profiles and thus allows a proper radiative transfer modeling
was introduced by \cite{Carroll07,Martinez08} and is based on a principal component analysis 
reconstruction technique. (see also Carroll et al. this proceeding).
Our code was applied to Stokes $I$ and $V$ observations of the K2 subgiant II Pegsi.
obtained with the SOFIN spectrograph at the Nordic Optical Telescope (La Palma).
The data were recorded in Summer 2004 and 2007 and cover 8 and 10 rotational phases respectively.
The original S/N of the data were 270 and after the application of our multi-line PCA reconstruction
(with 880 spectral lines) we could enhance the S/N of our target lines to 2000.
\begin{figure}
 \includegraphics[width=14cm]{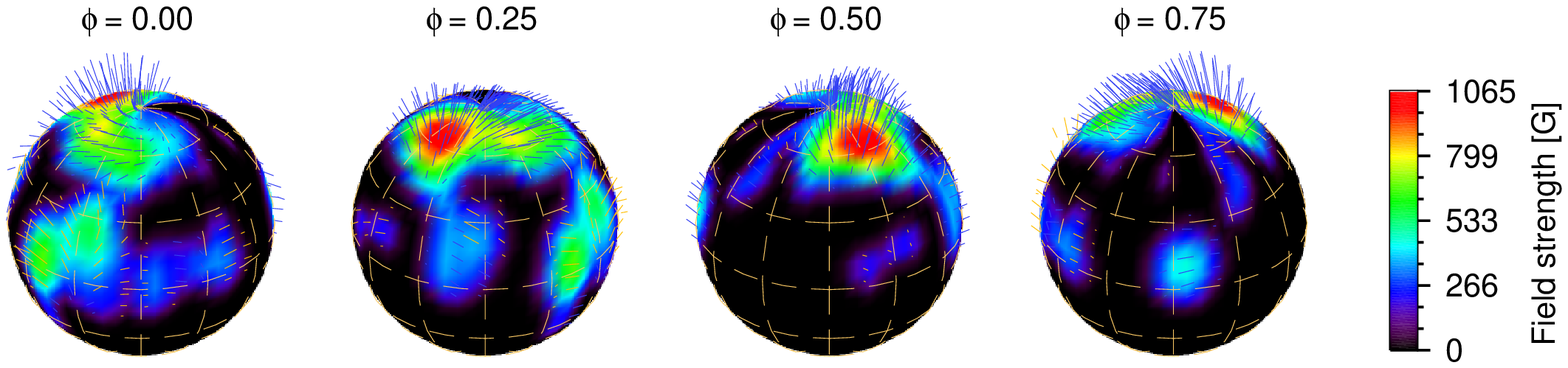}
 \includegraphics[width=14cm]{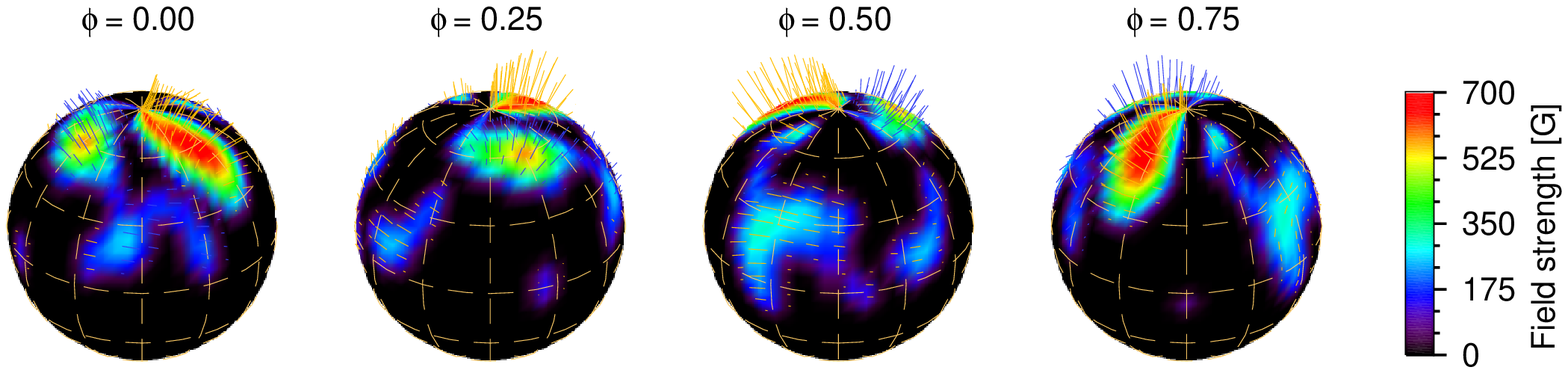}
  \caption{The reconstructed surface magnetic field of II Peg, 2004 (top) and 2007 (bottom). Bright and
  dark field lines indicate the polarities.}  
  \label{Fig:1}
\end{figure}
  
\section{The surface magnetic field of II Peg in 2004 and 2007}
\label{Sect:2}

From the reconstruction (Figure \ref{Fig:1}) we can readily identify that the magnetic field
of II Peg consist of a small number of large scale magnetic structures which
are predominantly located at high latitudes and arranged in active longitudes.
It is conspicuous that the field is mainly radially oriented an effect which we
attribute to our improved S/N ratio that allows our inversion to better constrain
the field orientation (see also Kochukhov et al. this proceeding).
A very interesting evolution can be seen from Figure \ref{Fig:1}, the field structure
has changed from a more or less coherent unipolar region in 2004 into
two separate field structures of different polarity. This characteristic 
change of the surface magnetic field topology resembles that of a flip-flop
event as found from temperature inversions \citep{Berdyugina07} 
and also found in theoretical modeling of an $\alpha-\Omega$ dynamo \citep{Elstner05}.
Although it is to early to confirm the presence of a real flip-flop event 
from our two year data, this work provides the first real 
indication of a magnetic reorganization on II Peg.

\end{document}